\documentclass{article}[12pt]
\usepackage{amsmath}
\usepackage{latexsym}
\usepackage{float}
\usepackage{amssymb}
\usepackage{graphicx}
\usepackage{epsfig}
\usepackage{cite}

\begin{document}

\title{Structure and diffusion in amorphous aluminium silicate: A molecular dynamics computer simulation}

\author{Anke Winkler$^{\text{1)}}$, J\"urgen Horbach$^{\text{1)}}$,
        Walter Kob$^{\text{2)}}$, and Kurt Binder$^{\text{1)}}$ \\[\baselineskip]%
    $^{\text{1)}}$\textit{Institut f\"ur Physik, Johannes Gutenberg--Universit\"at,} \\
                  \textit {D--55099 Mainz, Staudinger Weg 7, Germany}\\%
    $^{\text{2)}}$\textit{Laboratoire des Verres, Universit\'e Montpellier II,}\\
                  \textit{F--34095 Montpellier, France}}
\date{}

\maketitle

\begin{abstract}
The amorphous aluminium silicate (Al$_2$O$_3$)2(SiO$_2$) [AS2]
is investigated by means of large scale molecular dynamics computer
simulations.  We consider fully equilibrated melts in the temperature
range $6100$~K $ \ge T \ge 2300$~K as well as glass configurations
that were obtained from cooling runs from $T=2300$~K to $300$~K with a
cooling rate of about $10^{12}$~K/s.  Already at temperatures as high as
4000~K, most of the Al and Si atoms are four--fold coordinated by oxygen
atoms. Thus, the structure of AS2 is that of a disordered tetrahedral
network. The packing of AlO$_4$ tetrahedra is very different from that
of SiO$_4$ tetrahedra in that Al is involved with a relatively high
probability in small--membered rings and in triclusters in which an O
atom is surrounded by four cations.  We find as typical configurations
two--membered rings with two Al atoms in which the shared O atoms form
a tricluster. On larger length scales, the system shows a microphase
separation in which the Al--rich network structure percolates through
the SiO$_2$ network. The latter structure gives rise to a prepeak
in the static structure factor at a wavenumber $q=0.5$~\AA$^{-1}$.
The comparison of experimental X--ray data with the results from
the simulation shows a good agreement for the structure function. The
diffusion dynamics in AS2 is found to be much faster than in SiO$_2$. We
show that the self--diffusion constants for O and Al are very similar and
that they are by a factor of 2--3 larger than the one for Si.
\end{abstract}

\clearpage

\section{Introduction}
\label{sec1}

Amorphous mixtures of SiO$_2$ with other oxides such as Na$_2$O or
Al$_2$O$_3$ are of fundamental interest in geosciences~\cite{stebbins}
and glass technology~\cite{bach}. These systems exhibit many of the
physical phenomena that one encounters in multicomponent melts: E.g.~in
sodium silicates the sodium ions are much more mobile than the silicon
and oxygen atoms which leads to the property of ion conductance at
low temperature~\cite{ngai}.  Or in the mixture SiO$_2$--Al$_2$O$_3$
a miscibility gap emerges which is already present at 
temperatures slightly below 2000~K~\cite{hlavac83,rawson80,mcdowell69}.
Some recent computer simulations have shown that in order to understand
these properties of silicate melts one needs an accurate knowledge
of their microscopic structure~\cite{poole95}: For instance,
pure amorphous silica forms a network of corner--shared SiO$_4$
tetrahedra and it is known from Molecular Dynamics (MD) computer
simulations that diffusive motions are dominated by the existence
of defects such as SiO$_3$ and SiO$_5$ units in the network~(see,
e.g., Refs.~\cite{horb99_2,oligschleger99}).  Another example
is the sodium diffusion in sodium silicates: Molecular dynamics
simulation studies have demonstrated that the sodium diffusion can
be understood in terms of a motion through a channel network which is
embedded in the SiO$_2$ matrix~\cite{ingram,smith95,oviedo98,jund01},
and that the characteristic length scale of this channel network
is directly related to a prepeak in the static structure factor at
$q=0.95$~\AA$^{-1}$~\cite{horb02_1,horb03_1} the existence of which has
been verified in a recent neutron scattering experiment~\cite{meyer02}.
Despite this progress in our understanding of the microscopic structure
of these materials, many properties of silicates are still understood
only poorly. E.g.~the influence of a second network former, such as Al,
on the structural and dynamical properties, or on the phase diagram are
still not very clear. Therefore, the goal of the present paper is to
shed some light on this matter.

Aluminium silicate glasses have been investigated by means of
different experimental techniques such as Nuclear Magnetic Resonance
(NMR)~\cite{risbud87,sato91,poe92,meinhold93,schmuecker96,schmuecker97,schmuecker99},
IR and Raman spectroscopy, and X--ray
scattering~\cite{morikawa82,mcmillan82,okuno03}.  One of the main
issues in these experiments has been the analysis of the local
structure around the aluminium atoms. This is of special importance
for understanding the chemical ordering in aluminium silicates since
in principle the Al$^{3+}$ ions do need a different environment
of O$^{2-}$ ions than the Si$^{4+}$ ions in order to provide local
charge neutrality. Noteworthy are two peculiarities that have been
found in the experimental studies that distinguish the local oxygen
environment of an Al from that of a Si atom: In systems with a
high Al$_2$O$_3$ content such as mullite 3(Al$_2$O$_3$)2(SiO$_2$)
there seems to be a relatively large amount of five-- and six--fold
coordinated Al atoms in addition to AlO$_4$ units. And secondly,
NMR experiments found evidence for the existence of a high amount of
so--called triclusters, i.e.~structural units where an oxygen atom is
surrounded by three cations (whereby at least one of them is an aluminium
atom)~\cite{poe92,meinhold93,schmuecker96,schmuecker97,schmuecker99}.
The possibility of such triclusters has been very recently confirmed
in {\it ab initio}, molecular orbital calculations~\cite{kubicki02}.
The presence of triclusters has been also discussed in the context
of extensive viscosity measurements of Na$_2$O--Al$_2$O$_3$--SiO$_2$
liquids~\cite{toplis97_1,toplis97_2}.

As mentioned before, pure silica forms a network of corner--shared SiO$_4$
units and thus the O atoms are two--fold coordinated by Si atoms.
The appearance of triclusters in amorphous aluminium silicates is
an indication for the different local ordering of the different
cationic species Al and Si.  The difference in the local structure
around the Al atoms from that around the Si atoms is accompanied by
the tendency towards a metastable liquid--liquid phase separation
below $\approx 1900$~K for an Al$_2$O$_3$ content between about 10 to
50~mol\%~\cite{hlavac83,rawson80,mcdowell69}. The system we
consider in the following is approximately in the centre of the
demixing region: By means of MD simulations we investigate an aluminium
silicate melt with 33~mol\% Al$_2$O$_3$, i.e.~(Al$_2$O$_3$)2(SiO$_2$),
and in the following we will denote this system by AS2.

It is one of the merits of MD simulations that one can study at the same
time the structure on local and medium length scales in conjunction
with dynamic properties. The duration of the MD runs is of course
restricted to time scales in the ns range and so we are not able
to approach the experimentally expected liquid--liquid coexistence
line very closely. However, as we demonstrate in the following, the
simulations do shed light onto precursors of phase separation in the
microscopic structure.

The rest of the paper is organized as follows: In the next section we
give the details of the model and the simulation procedure. Then we present
the results for the structure and diffusion in AS2, and finally we
summarize and discuss them in the last section. 

\section{Model and details of the simulation}
\label{sec2}

A potential that has often been used recently to study the properties
of amorphous silica is the so--called BKS potential~\cite{bks} that
has been developed by van Beest, Kramer, and van Santen by means of {\it ab
initio} calculations. Although it is a simple pair potential, it has been shown
to reproduce very well many static and dynamic properties of amorphous 
silica~\cite{vollmayr96,barrat97,hemmati98,horb99_1,horb99_2,horb01_1,horb01_2,horb01_3,roder01,mischler02,jund99,benoit00,lacks00,saika01,shell02}.
An extension of the BKS potential that allows to consider also
mixtures of silica with other oxides such as Na$_2$O and Al$_2$O$_3$
was proposed by Kramer {\it et al.}~\cite{kramer}.  As demonstrated
recently~\cite{horb99_3,horb01_4,horb02_1,horb02_2,horb03_1,jund01,ispas01,winkler03},
this potential gives also a quite realistic description of the static
and dynamic properties of sodium silicates.

In this work we use the latter potential to investigate the aluminium
silicate melt (Al$_2$O$_3$)2(SiO$_2$) [AS2].  The functional form of
the potential is as follows:
\begin{equation}
\varphi_{\alpha\beta}(r)=
\frac{q_\alpha q_\beta e^2}{r}
+ A_{\alpha\beta}\exp({-B_{\alpha\beta}r}) - \frac{C_{\alpha\beta}}{r^6} 
\,,
\label{eq1}
\end{equation}
where $\alpha ,\beta \in \{\mathrm{Si,Al,O}\}$.  Here $r$
is the distance between an ion of species $\alpha$ and
an ion of species $\beta$. The values of the parameters
$\{A_{\alpha\beta},B_{\alpha\beta},C_{\alpha\beta}\}$ that were
calculated by {\it ab initio} methods are $A_{\rm SiO}=18003.7572$~eV,
$A_{\rm AlO}=8566.5434$~eV, $A_{\rm OO}=1388.7730$~eV, $B_{\rm
SiO}=4.87318$~\AA$^{-1}$, $B_{\rm AlO}=4.66222$~\AA$^{-1}$, $B_{\rm
OO}=2.76$~\AA$^{-1}$, $C_{\rm SiO}=133.5381$~eV\AA$^{6}$, $C_{\rm
AlO}=73.0913$~eV\AA$^{6}$, and $C_{\rm OO}=175.0$~eV\AA$^{6}$ (for the
Si--Si, Si--Al and Al--Al interactions the latter parameters are all
set to zero)~\cite{kramer}.

From Eq.~(\ref{eq1}) it becomes obvious that at small distances the
potential between the Al (or Si) and the O atoms goes to minus infinity
(since the coefficients $C_{\alpha\beta}$ are positive), i.e. it becomes
unphysical. Therefore we have modified the potential at short distances
by substituting it with a simple parabola:
\begin{equation}
  \varphi_{\rm AlO} = - 19.9508 \; {\rm eV}  + 15.5 \; \frac{{\rm eV}}{{\mbox{\AA}}^2}
                     \left( r - 1.16576 \; {\mbox{\AA}} \right)^2
                    \quad \quad r \le 1.1658 \; {\mbox{\AA}}
  \  .
  \label{eq4}
\end{equation}
We emphasize however, that this change hardly affects the properties of
the system since at intermediate and low temperatures the probability
of finding a distance below 1.17~\AA~between an Al and an O particle is
very small. Note that we use similar modifications as the one given by
Eq.~(\ref{eq4}) also for the Si--O and O--O interactions. Details can
be found in Ref.~\cite{vollmayr96}.

\begin{figure}
\psfig{file=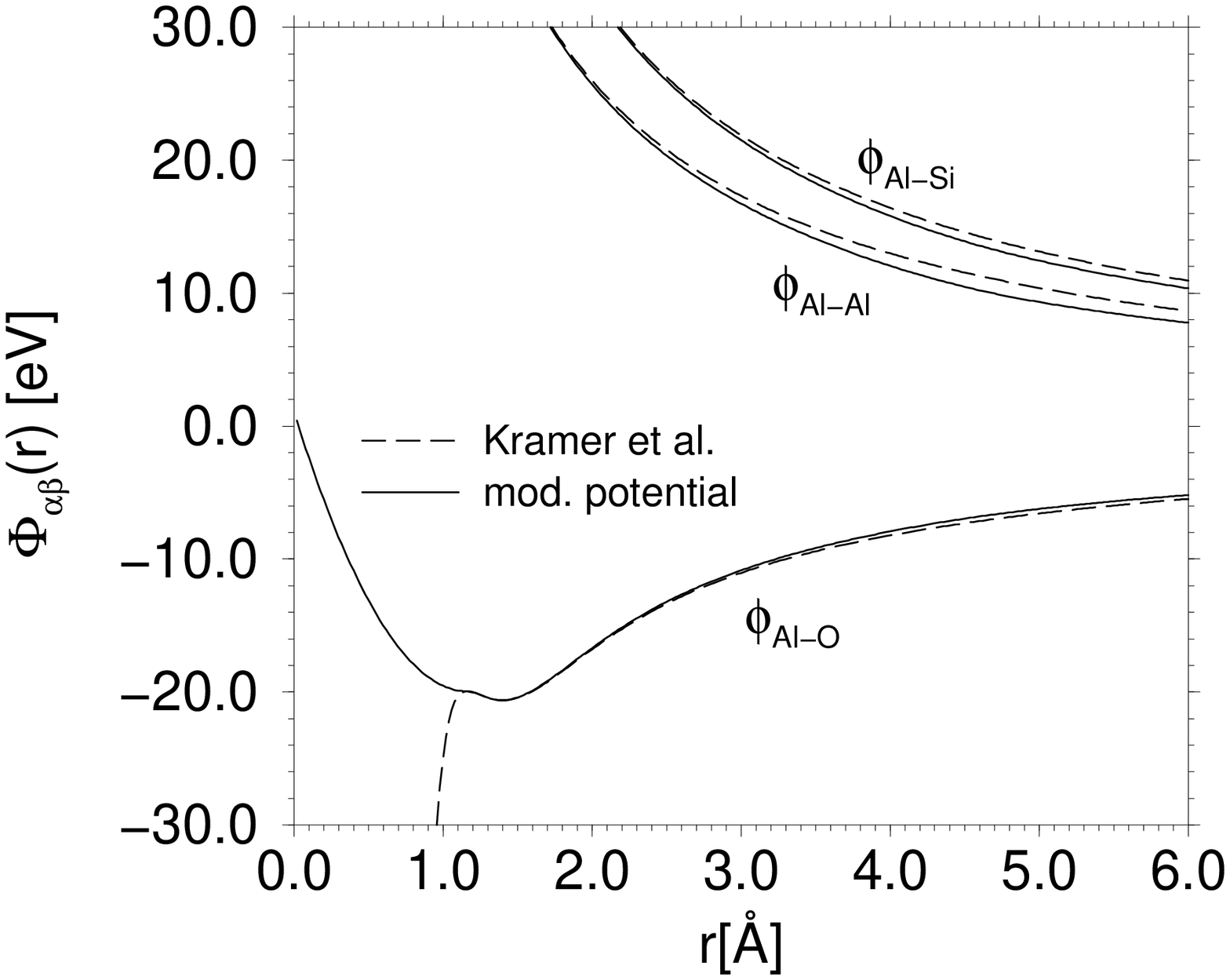,height=10cm}
\caption{Al--Si, Al--Al and Al--O interaction potentials as proposed by 
Kramer {\it et al.}~\cite{kramer} (dashed lines) and in the modified version
(solid lines) according to Eqs.~(\ref{eq2})--(\ref{eq4}).}
\label{fig1}
\end{figure}

In the long--ranged Coulomb--part the charges $q_{\alpha} e$ ($e$:
charge of an electron) are not the bare ionic charges of ions of type
$\alpha$ but are considered to be effective charges. Unfortunately in
Ref.~\cite{kramer}, the $q_{\alpha}$'s were fixed such that systems
like AS2 are not neutral: $q_\mathrm{O}=-1.2$, $q_\mathrm{Si}=2.4$,
and $q_\mathrm{Al}=1.9$. Only with the additional component phosphorus
the system recovers charge neutrality according to the parameter sets in
Ref.~\cite{kramer}. The same problem arises for sodium silicates if one
uses the Kramer potential. In order to overcome this problem we follow
here the same strategy that we have already applied in the case of the
potential for the sodium silicates~\cite{horb99_3,horb02_1,horb02_2}:
We set the charge of aluminium to $q_\mathrm{Al} e =1.8 e$ such that a
stoichiometric aluminium silicate system as AS2 is neutral.  Then we add
a short--ranged potential that compensates for this change at {\it short}
distances. The final form of the potential is given by
\begin{equation}
  \Phi_{\alpha \beta} = \varphi_{\alpha \beta}
         + \frac{\tilde{q}_{\alpha} \tilde{q}_{\beta} e^2}{r}
          [ 1 - ( 1- \delta_{\alpha {\rm Al}} ) ( 1- \delta_{\beta {\rm Al}} ) ]
          \Theta ( r_{\rm c} - r ) \quad ,
  \label{eq2}
\end{equation}
with $\Theta$ being the Heaviside function, 
with $\tilde{q}_{\rm Si}=2.4$, $\tilde{q}_{\rm O}=-1.2$, and
\begin{equation}
\tilde{q}_\mathrm{Al}(r)=
q_\mathrm{Al} \left( 1+ 
                    \ln\left[C
              \frac{(r_{\rm c}-r)^2}{1 {\rm \AA}^2 +(r_{\rm c}-r)^2}+1\right]
              \exp\left(-\frac{d}{(r-r_{\rm c})^2}\right)
 \right)  \ .
\label{eq3}
\end{equation}
Here the parameters $r_{\rm c}=6.0$~\AA, $C =0.0653$, and $d=2.0$~\AA$^2$
are chosen. These parameters guarantee that the potential given by
Eq.~(\ref{eq2}) is indeed very similar to the original one as illustrated
in Fig.~\ref{fig1}: For $r < 2.2$~\AA~the modified potential is almost
the same as the original one. Finally we mention that the presence of the
exponential term in Eq.~(\ref{eq3}) makes that the charge $q_\mathrm{Al}$
is a smooth function of $r$.

The equations of motion were integrated with the velocity form of
the Verlet algorithm using a time step of 1.6~fs. The Coulomb forces
were calculated by the standard Ewald summation technique.  At each
temperature we equilibrated the system first in the NVT ensemble by
coupling it to a stochastic heat bath. Thereby, the equilibration time
exceeded the structural relaxation time for the slowest component
(i.e. silicon), i.e.~dynamic density--density correlation functions
for silicon at the wave--vector $q=1.7$~\AA$^{-1}$ (corresponding to
the characteristic length scale of the tetrahedral network, see below)
have decayed to zero in that time~\cite{winkler02}.  After equilibration
the heat bath was switched off, and we started production runs in the
microcanonical ensemble.  During all runs the density was fixed to the
experimental value at 300 K, $\rho=2.60$~g/cm$^3$~\cite{mazurin}. The
number of particles was 1408 ($N_{\rm Si}=256$, $N_{\rm Al}=256$, $N_{\rm
O}=898$) and hence the linear dimension of the cubic simulation box was
$L=26.347$~\AA. The temperatures that we have investigated are 6100~K,
4700~K, 4000~K, 3580~K, 3250~K, 3000~K, 2750~K, 2600~K, 2480~K, 2380~K,
and 2300~K.  The production runs at the lowest temperature lasted over
6.9~ns real time corresponding to 4.2~million time steps. In addition we
did cooling runs from 2300~K to 0~K with a constant cooling rate $\gamma
= 1.42\cdot 10^{12}$~K/s.  At each temperature we did five completely
independent runs in order to improve the statistics.

\section{Results}
\label{sec3}

In this section we present the results of our simulations. First,
we study in detail the local structure of the network in AS2 and show
then the consequences of the local chemical ordering for length scales
that go beyond distances of nearest and next--nearest atoms. Finally,
we discuss the behavior of the self--diffusion constants.

\begin{figure}
\psfig{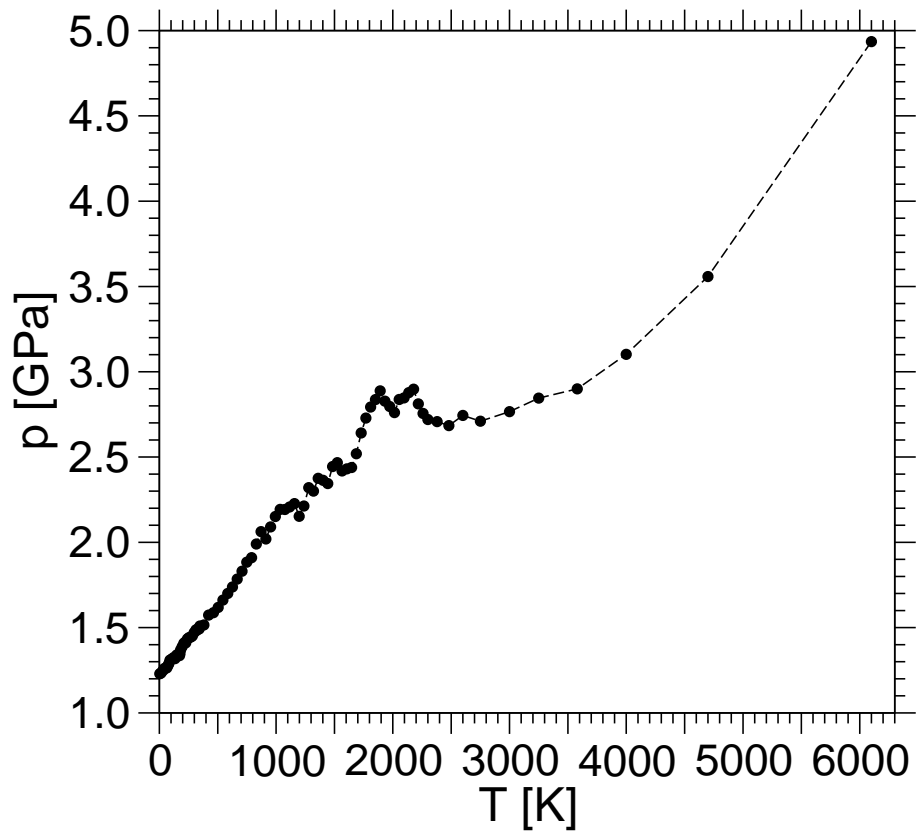}
\caption{The pressure as a function of temperature. Note that the points for
$T < 2300$~K are sampled during runs where the system was cooled down from
2300~K to 0~K with a cooling rate of $\gamma = 1.42 \cdot 10^{12}$~K/s.}
\label{fig2}
\end{figure}

Before we start the discussion of structural quantities we present in
Fig.~\ref{fig2} the pressure $p$ as a function of temperature.  From this
graph we recognize that in the temperature range considered the pressure
varies between 1 and 5~GPa. This strong variation is related to the fact
that we have made our simulation at constant volume. It is, however,
reassuring that at ambient temperature the pressure is not exceedingly
high, thus showing that the potential is quite reliable with respect
to the pressure. Furthermore we point out that it is unlikely that for
a system like AS2 the structure changes significantly in the pressure
range that we have here. Therefore we expect that the results presented
here are very similar to the ones that one would get in a constant
pressure simulation.

It is remarkable that $p(T)$ exhibits a local minimum around $T=2500$~K
which would correspond to a density maximum in a constant pressure
simulation. So our model predicts for AS2 a density anomaly which is a
well--known feature in amorphous silica~\cite{brueckner70}.  Note that
for $T < 2300$~K the system is no longer in equilibrium but in the
glass phase. From the fact that in this temperature range $p(T)$ shows
basically a linear variation with temperature, we thus can conclude that
anharmonic effects are not important.

\begin{figure}
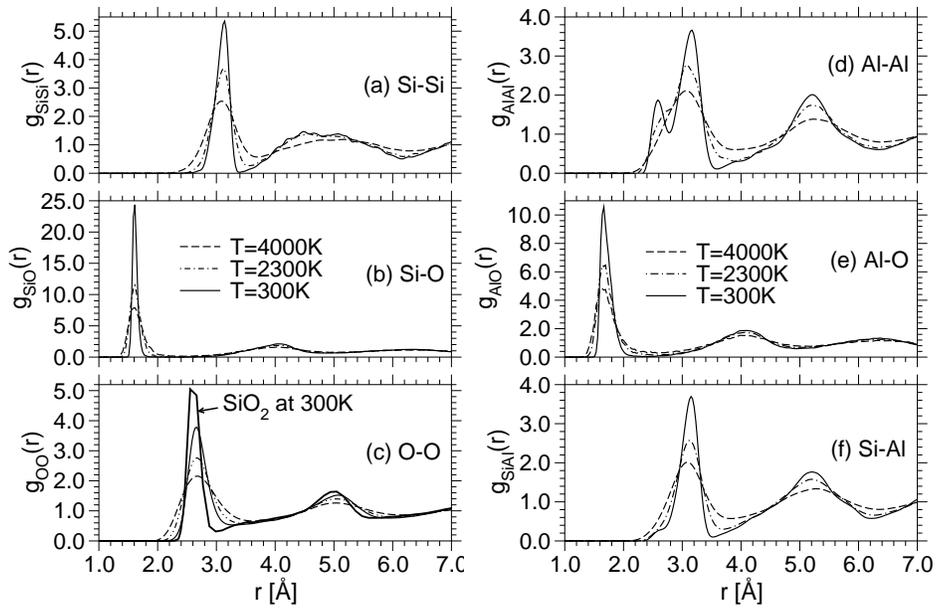

\psfig{file=fig3abc.eps,height=8cm}
\psfig{file=fig3def.eps,height=8cm}
\caption{Partial pair correlation functions $g_{\alpha \beta}(r)$ for the
temperatures $T=4000$~K, 2300~K and 300~K. a) $g_{\rm SiSi}(r)$, b) $g_{\rm SiO}(r)$, 
c) $g_{\rm OO}(r)$, d) $g_{\rm AlAl}(r)$, e) $g_{\rm AlO}(r)$, f) $g_{\rm SiAl}(r)$. 
The curve in c) for SiO$_2$ at $T=300$~K is taken from Ref.~\cite{horb99_2}.}
\label{fig3}
\end{figure}

Quantities that are well suited to characterize the local structure of
atomic systems are the partial pair correlation functions $g_{\alpha
\beta} (r)$ which are proportional to the probability of finding a
particle of type $\beta$ at a distance $r$ from a particle of type
$\alpha$. The definition of $g_{\alpha \beta} (r)$ can be found in
standard textbooks~\cite{hansen}. The six different $g_{\alpha \beta}
(r)$ of our system are shown in Fig.~\ref{fig3} for the three temperatures
$T=4000$~K, 2300~K, and 300~K. The sharp first peak around $r=1.605$~\AA\
in $g_{\rm SiO} (r)$ reflects the strong covalent nature of the Si--O
bond.  We also can note that in the first minimum of $g_{\rm SiO}
(r)$, which is around 2.3~\AA, the function is basically zero even at
4000~K. Thus this feature allows for a natural definition of nearest
neighbors of a silicon atom, and below we will make use of this fact.
Moreover, there is a gap between 2~\AA\ and 3~\AA\ which is due to the
chemical ordering in the network of AS2: In between a silicon atom and a
second nearest oxygen neighbor there must be always another silicon (or
aluminium) atom.  The functions $g_{\rm AlO} (r)$ look very similar to
$g_{\rm SiO} (r)$ at the corresponding temperature, but the former
are less pronounced in that the first peak is slightly broader and it
has a smaller amplitude. Furthermore, the position of the first peak in
$g_{\rm AlO} (r)$ is now at the slightly higher value $r=1.66$~\AA. This
is in agreement with experiments and {\it ab initio} simulations of
similar systems, although slightly higher values between 1.71~\AA\
and 1.77~\AA\, have been reported~\cite{benoit01,shannon76}.

Significant differences are found between $g_{\rm AlAl} (r)$ and
$g_{\rm SiSi} (r)$: At 300~K the first peak in $g_{\rm AlAl} (r)$
splits up in two peaks at $r_1=2.59$~\AA\ and $r_2=3.16$~\AA\ whereas
in $g_{\rm SiSi} (r)$ one finds only a single peak at about 3.12~\AA,
i.e.~a value very similar to $r_2$.  Also in $g_{\rm SiAl} (r)$ one finds
a shoulder around $r_1$. We will see below that the feature around $r_1$
is due to the presence of two--membered rings. Fig.~\ref{fig3}c shows a
comparison of $g_{\rm OO} (r)$ for AS2 with that for SiO$_2$ at $T=300$~K
(the latter was taken from a recent simulation study, for details see
Ref.~\cite{horb99_2}). The main difference is that the first peak in
the function for AS2 is shifted to larger distances which stems from
the fact that the length of an Al--O bond is slightly larger than the
length of a Si--O bond.

In pure SiO$_2$ a disordered tetrahedral network is formed such that
a silicon atom sits in the centre of each tetrahedron, whereby the
oxygen atoms at the four corners of this tetrahedron are shared by the
silicon atoms of the two neighboring tetrahedra (and thus each oxygen
atom is two--fold coordinated by silicon atoms). It was shown in a MD
simulation of a SiO$_2$ model~\cite{horb99_2}) that this local structure
is essentially formed at temperatures as high as 3000~K since even
at this temperature the percentage of defects (such as a Si atom that
is five--fold coordinated by O atoms or an O atom that is three--fold
coordinated by Si atoms) is smaller than 5~\%.

\begin{figure}
\psfig{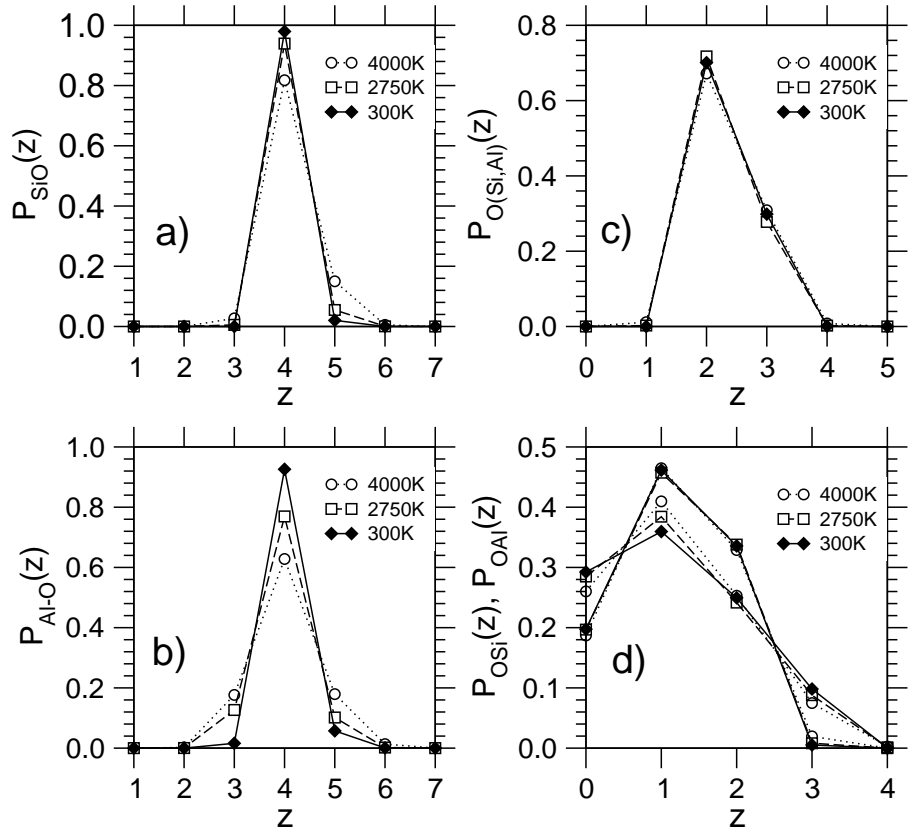}
\caption{Distribution of coordination numbers $P_{\alpha \beta}(z)$ for the
temperatures $T=4000$~K, 2300~K and 300~K, a) $P_{\rm SiO}(z)$, b) $P_{\rm AlO}(z)$, 
c) $P_{\rm O(Si,Al)}(z)$, d) $P_{\rm OSi}(z)$ and $P_{\rm OAl}(z)$. The three
curves that at $z=2$ have the largest values correspond to $P_{\rm OSi}(z)$.}
\label{fig4}
\end{figure}

We demonstrate now that the structure of our AS2 model is far from a
perfect tetrahedral network even at very low temperatures.  To this end, we
consider coordination number distributions $P_{\alpha \beta}(z)$ which
give the probability that a particle of type $\alpha$ is surrounded by
exactly $z$ neighbors of type $\beta$ within a distance $r \le r_{\rm
min}^{\alpha \beta}$ (where $r_{\rm min}^{\alpha \beta}$ corresponds
to the first minimum in $g_{\alpha \beta}(r)$).  Fig.~\ref{fig4} shows
$P_{\alpha \beta}(z)$ for Si--O, Al--O, O--(Si,Al), O--Si, and O--Al
correlations (O--(Si,Al) means that one does not distinguish between Si
and Al atoms) at the three temperatures 4000~K, 2750~K, and 300~K. The
values used for $r_{\rm min}^{\alpha \beta}$ are $r_{\rm min}^{\rm
SiO}=2.33$~\AA, $r_{\rm min}^{\rm AlO}=2.54$~\AA\ for $T=4000$~K,
$r_{\rm min}^{\rm SiO}=2.25$~\AA, $r_{\rm min}^{\rm AlO}=2.42$~\AA\
for $T=2750$~K, and $r_{\rm min}^{\rm SiO}=2.20$~\AA, $r_{\rm min}^{\rm
AlO}=2.40$~\AA\ for $T=300$~K.  As we can infer from $P_{\rm SiO}$
and $P_{\rm AlO}$, at low temperatures most of the Si and Al atoms are
four--fold coordinated by O atoms and thus also Al is integrated in the
tetrahedral network structure.  Note that also a recent MD simulation of
a realistic model of a pure Al$_2$O$_3$ melt finds a structure at low
temperatures in which most of the Al atoms are four--fold coordinated
by O atoms~\cite{guti02}.

That the packing of the AlO$_4$ tetrahedra is nevertheless different
from that of the SiO$_4$ tetrahedra is demonstrated in Figs.~\ref{fig4}c
and \ref{fig4}d.  Only 70\% of the O atoms are two--fold coordinated but
around 30\% of the O atoms are three--fold coordinated by (Si,Al) atoms,
thus forming the so--called triclusters that have been identified in NMR
experiments.  It is remarkable that the percentage of such triclusters is
nearly independent of temperature in our simulations.  The distributions
$P_{\rm OSi}$ and $P_{\rm OAl}$ show that the probability that an O atoms
is three--fold coordinated by silicon atoms is very low (essentially zero
at 300~K), whereas the probability for triclusters with three Al atoms is
relatively high and increases even slightly with decreasing temperature. A
closer inspection at $T=300$~K shows the following cation composition
of the triclusters: There are 1.4\% with three Si, 12.4\% with two Si
and one Al, 47.3\% with one Si and two Al, and 38.9\% with three Al.
So most triclusters contain one Si and two Al atoms or three Al atoms.

\begin{figure}
\psfig{file=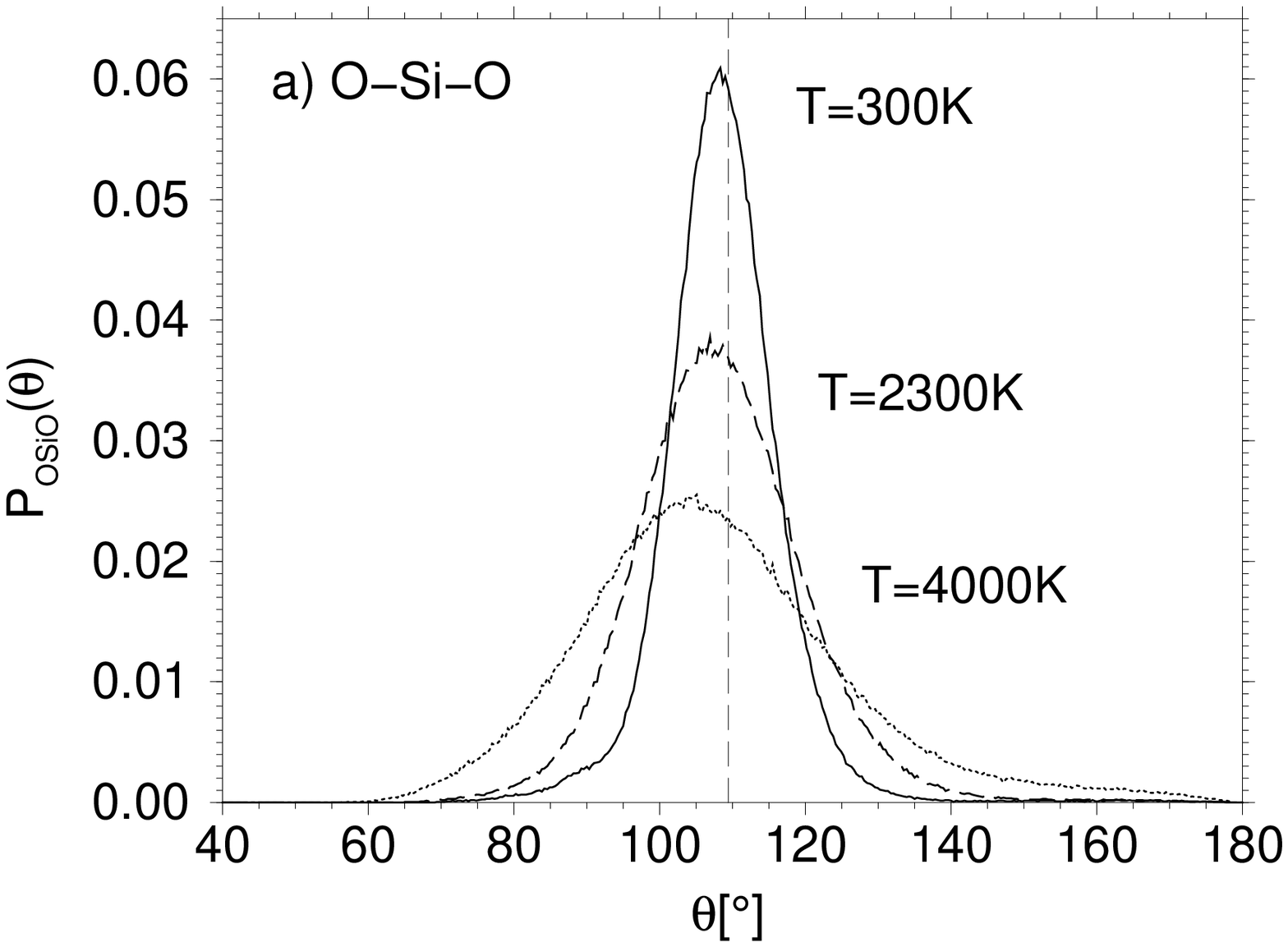,height=9cm}
\psfig{file=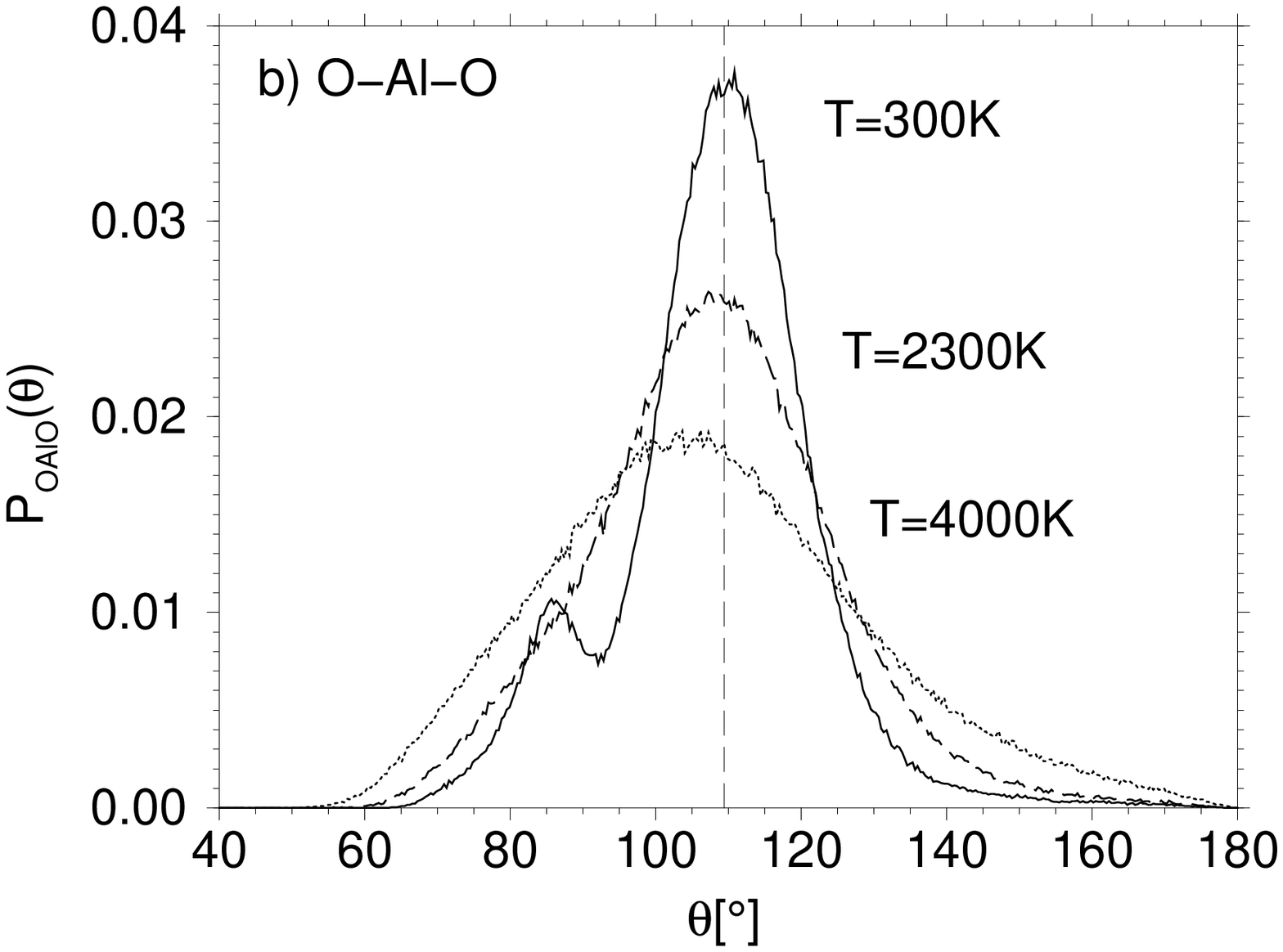,height=9cm}
\caption{Distributions of angles $\theta$ for the
temperatures $T=4000$~K, 2300~K and 300~K, a) $P_{\rm OSiO}(\theta)$, 
b) $P_{\rm OAlO}(\theta)$. The vertical lines correspond to the ideal
tetrahedron angle of 109.47$^{\circ}$.}
\label{fig5}
\end{figure}

We have seen that at low temperatures most of the Al atoms are four--fold
coordinated by oxygens. On the other hand there are two different
characteristic length scales $r_1$ and $r_2$ for the distance between
nearest Al neighbors.  These two length scales should be also reflected
in the geometry of the AlO$_4$ tetrahedra: Two connected tetrahedra
for which the Al atoms are at a distance $r_1$ from each other may
have a different geometry from two connected tetrahedra where the two
aluminium atoms in the centres are at distance around $r_2$. Appropriate
quantities to study the geometry of the tetrahedra are $P_{\alpha \beta
\gamma}(\theta)$, the distribution for the O--Si--O and O--Al--O angles,
which are shown in Fig.~\ref{fig5}. In $P_{\rm OSiO}(\theta)$ a single
peak is observed which, by decreasing the temperature, becomes sharper
and the location of the maximum moves to larger angles. At $T=300$~K the
maximum is at $\theta=108.2^{\circ}$ which is close to the value for an
ideal tetrahedron, $\theta=109.47^{\circ}$ (vertical lines). The behavior
of $P_{\rm OAlO}(\theta)$ is very different in that one finds at $T=300$~K
two peaks: $\theta_1=85.8^{\circ}$ and $\theta_2=109.8^{\circ}$. Such
a bimodal distribution has been found also in computer simulations
of free silica surfaces~\cite{roder01,mischler02} where one obtains
two peaks in $P_{\rm OSiO}(\theta)$ for the surface region at similar
values for $\theta_1$ and $\theta_2$. And also in this case the bimodal
distribution of angles is accompanied by two ``nearest neighbor peaks''
in $g_{\rm SiSi}(r)$. The authors of Refs.~\cite{roder01,mischler02}
have explained these features at free silica surfaces by the presence of
two--membered rings, i.e.~structural units where two tetrahedra share
two oxygen atoms. The two tetrahedra that form the two--membered rings
are deformed such that the distance between the two cations is at $r_1$
and the O--cation--O angles are at $\theta_1$. (Here the two oxygens are
of course the shared O atoms.) There is one major difference between
the two--membered rings of free silica surfaces and the ones in AS2:
In the former case they become very rare at low temperatures whereas in
the case of AS2 their occurrence increases slightly with temperature.

\begin{figure}
\psfig{file=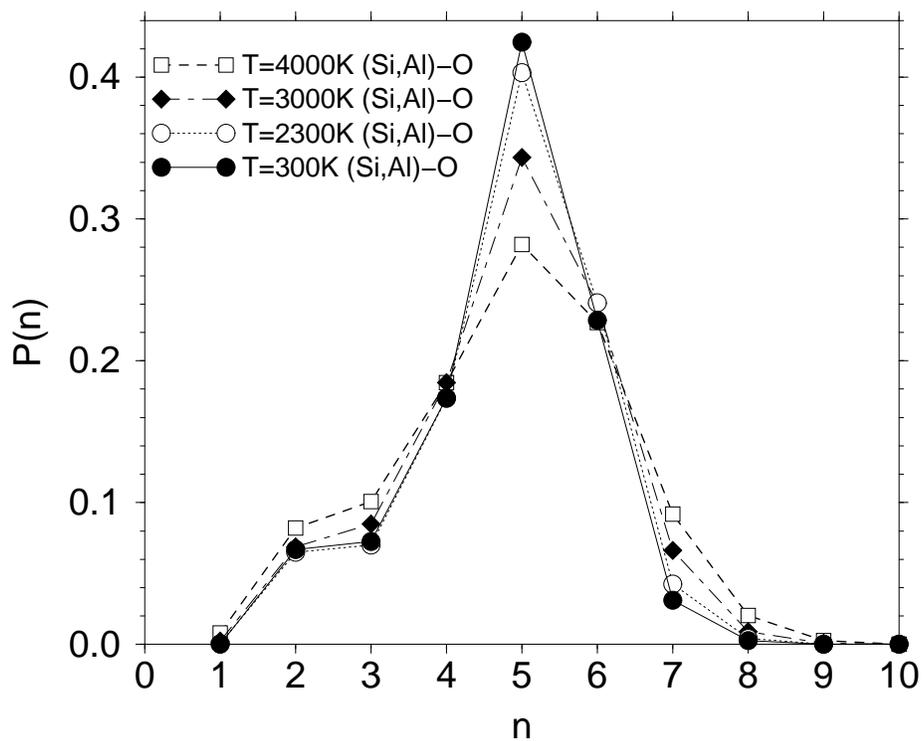,height=10cm}
\caption{Distribution of rings $P(n)$ for the temperatures $T=4000$~K, 
3000~K, 2300~K and 300~K. See text for the definition of a ring.}
\label{fig6}
\end{figure}

Apart from two--membered rings it is also straightforward to define rings
of any size $n$ in the network: One selects any cation (i.e.~Si or Al)
and two of its O neighbors. There are several possible paths how one
can move through the network structure from one cation--O pair to the
next one such that one starts from the pair with the first O atom and
ends at a cation--O pair with the second O atom.  The shortest possible
of such paths are called rings and the length of a ring is the number
of cation--O pair that it contains. In Fig.~\ref{fig6} the ring size
distribution $P(n)$ is shown for three different temperatures. Whereas
$P(n)$ changes significantly from 4000~K to 2300~K, only minor changes
are observed from 2300~K to 300~K. However, we recognize essentially
the same qualitative behavior at the three temperatures: most probable are
rings with a size of $n=5$ and there is a relatively large contribution of rings
with $n=2$ and $n=3$. Note that the location of the maximum at $n=5$ is
different from that which one expects for pure silica at low temperatures
where MD simulations have found the maximum at $n=6$ (corresponding to the
length of the rings in $\beta$--cristobalite)~\cite{vollmayr96,rino93}.

We have also determined the composition of small--membered rings at
$T=300$~K yielding for the rings with $n=2$ that 72.8\% consist of two
Al atoms, 25.0\% contain one Al and one Si atom, and there are 2.2\%
with two Si atoms.  For the three--membered rings we obtain the following
numbers: 36.5\% with three Al atoms, 45.3\% with two Al and one Si atom,
15.5\% with one Al and two Si atoms, and 2.7\% with three Si atoms.

\begin{figure}
\psfig{file=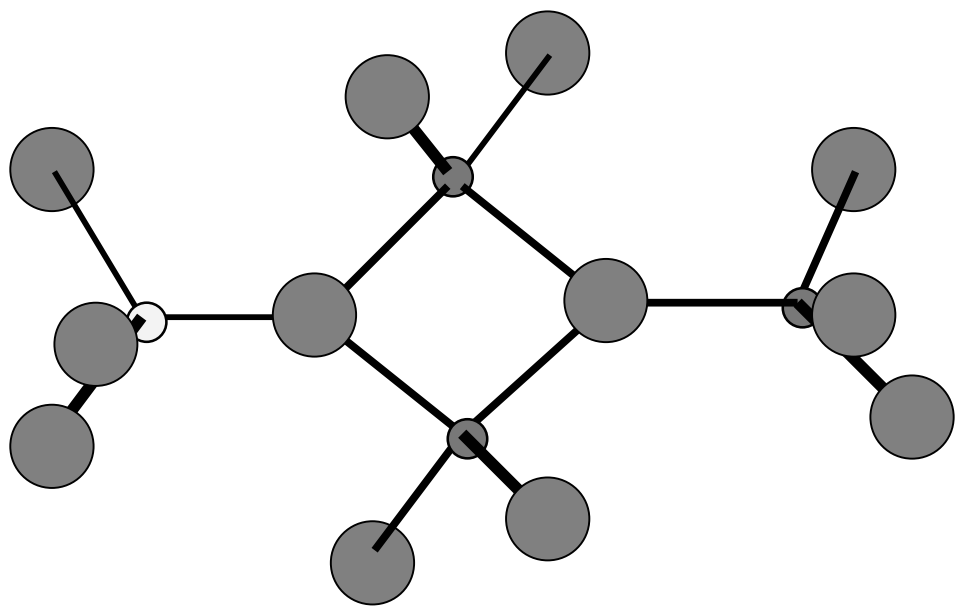,height=8cm}
\caption{Schematic picture of a typical local configuration in AS2. O atoms
are the big grey spheres, Al atoms are the small grey spheres, and the light small
sphere is a Si atom. The black lines between the atoms symbolize covalent
Al--O and Si--O bonds. The picture shows a two--membered Al--O ring whereby
the O atoms in this ring form triclusters.}
\label{fig7}
\end{figure}

One may ask whether the relatively high probability of two--membered
rings in our AS model is related to the occurrence of the aforementioned
triclusters. And indeed we have extracted from our data that at
$T=300$~K 96\% percent of the oxygen atoms that are involved in a
tricluster are also a member of a ring with $n=2$. And this holds also
the other way round: more than 95\% of the O atoms that are members in
any two--membered ring are three--fold coordinated by cations and form
thus a tricluster.  So we find that the appearance of triclusters in our
AS2 model is accompanied by the presence of two--membered rings which
to our knowledge is a connection that has not been considered yet in
the analysis of experiments. A typical local configuration with
a two--membered Al--O ring and two triclusters is illustrated by a
schematic picture in Fig.~\ref{fig7}.

\begin{figure}
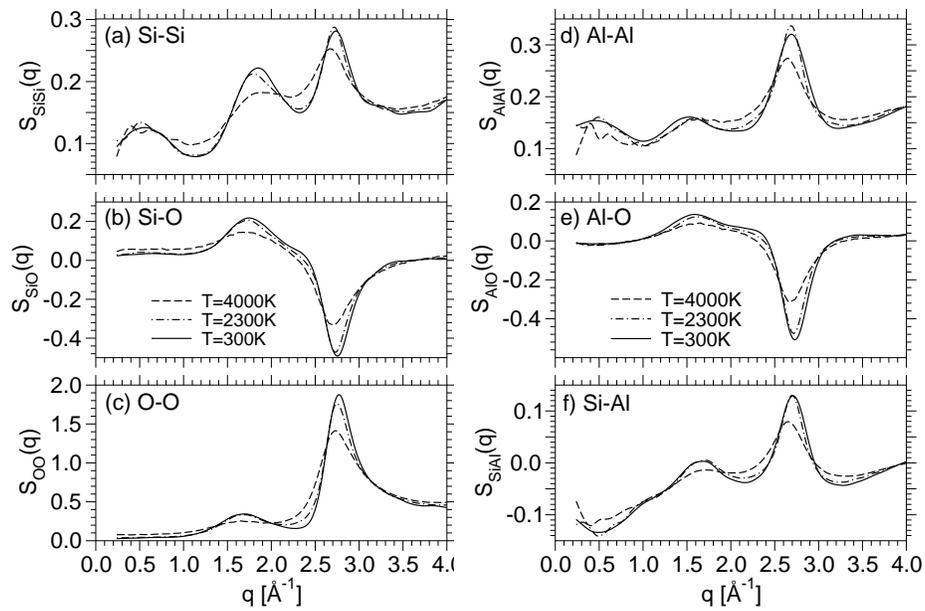

\psfig{file=fig8abc.eps,height=8cm}
\psfig{file=fig8def.eps,height=8cm}
\caption{Partial static structure factors $S_{\alpha \beta}(q)$ for the
temperatures $T=4000$~K, 2300~K and 300~K. a) $S_{\rm SiSi}(q)$, b) $S_{\rm SiO}(q)$, 
c) $S_{\rm OO}(q)$, d) $S_{\rm AlAl}(q)$, e) $S_{\rm AlO}(q)$, f) $S_{\rm SiAl}(q)$.} 
\label{fig8}
\end{figure}

The results discussed so far have shown that the aluminium atoms do form
tetrahedral units with oxygens but that the packing of these AlO$_4$
units is very different from the one of the SiO$_4$ units. We now want
to study whether the different local order around Al and Si atoms leads
to structural features that are present on larger length scales. To
this end, we consider partial static structure factors which are the
Fourier transforms of the corresponding pair correlation functions.
They are correlation functions of the number densities
\begin{equation}
    \rho_{\alpha}({\bf q}) =
   \sum_{k=1}^{N_\alpha} \exp( i {\bf q} \cdot {\bf r}_k )  \quad \quad \quad 
   \alpha \in \{{\rm Si, Al, O}\}
\label{eq5a}
\end{equation}
(depending on wave--vector ${\bf q}$) and can be defined as follows~\cite{hansen}:
\begin{eqnarray}
  S_{\alpha \beta}(q) & = &
   \frac{1}{N} \left< \rho_\alpha({\bf q}) \rho_\beta (-{\bf q}) \right>  \nonumber \\
  & = &  \frac{f_{\alpha\beta}}{N}
    \sum_{k=1}^{N_{\alpha}}  \sum_{l=1}^{N_{\beta}}
    \langle \exp(i {\bf q} \cdot ({\bf r}_k-{\bf r}_l )) \rangle \ ,
\label{eq5}
\end{eqnarray}
with $N$ being the total number of particles, $q$ is the absolute value
of the wave--vector ${\bf q}$, and $f_{\alpha\beta}$ is equal to $0.5$
if $\alpha\neq\beta$ and to $1.0$ if $\alpha = \beta$. Fig.~\ref{fig8}
shows $S_{\alpha \beta}(q)$ at the three temperatures $T=4000$~K, 2300~K,
and 300~K. Similar to the results found for the radial distribution
functions, the temperature dependence of $S_{\alpha \beta}(q)$
is relatively weak in that the different peaks that are found at low
temperatures are already present at a temperature as high as 4000~K. For
$q>2.3$~\AA$^{-1}$ the partial structure factors reflect length scales
of nearest neighbors and their location corresponds approximately to the
period of oscillations in the $g_{\alpha \beta}(r)$.  But we are now more
interested in the features in $S_{\alpha \beta}(q)$ at small $q$. The
peaks at $1.6$--$1.7$~\AA$^{-1}$ are due to the order that arises from
the tetrahedral network structure, i.e.~from repeated AlO$_4$ and SiO$_4$
units. And indeed the length $2 \pi/(1.7\; {\mbox{\AA}}^{-1})=3.7$~\AA\
corresponds approximately to the spatial extent of two connected
tetrahedra.  Note that a prepeak around $q=1.7$~\AA$^{-1}$ in the
static structure factor is also found in pure silica and in many other
materials that form tetrahedral networks (e.g. see Ref.~\cite{vollmayr96}
and references therein).

But from Fig.~\ref{fig8} we recognize that there is also an additional
prepeak at $q=0.5$~\AA$^{-1}$ in the $S_{\alpha \beta}(q)$ for the Si--Si,
Al--Al, and Si--Al correlations. One possible explanation for this peak is 
the following one: As we have seen in our analysis of the local structure
before, the local packing of AlO$_4$ tetrahedra is significantly different
from that of the SiO$_4$ tetrahedra. This may lead to a structure where an
AlO$_4$ tetrahedron prefers to be surrounded by other AlO$_4$ tetrahedra
and thus Al rich regions are formed. So a structure is created where the
Al atoms are not homogeneously distributed on the relatively large length
scale $l \approx 2 \pi/ (0.5\; {\mbox{\AA}}^{-1}) \approx 12.6$~\AA. If
one considers only the Al atoms for instance, voids are formed with the
spatial extent given by $l$ in the regions where the Si atoms sit. The
same holds of course true if one considers only the Si atoms. In the
$S_{\alpha \beta}(q)$ in which the O atoms are involved, almost no peak
at $0.5$~\AA$^{-1}$ is seen because the oxygen atoms are essentially
homogeneously distributed in the system on the length scale $l$ since
they are both nearest neighbors of Si and Al atoms (the difference in
the chemical ordering of the O atoms around the Al atoms from that around
the Si atoms is only weakly pronounced on the length scale $l$).

\begin{figure}
\psfig{file=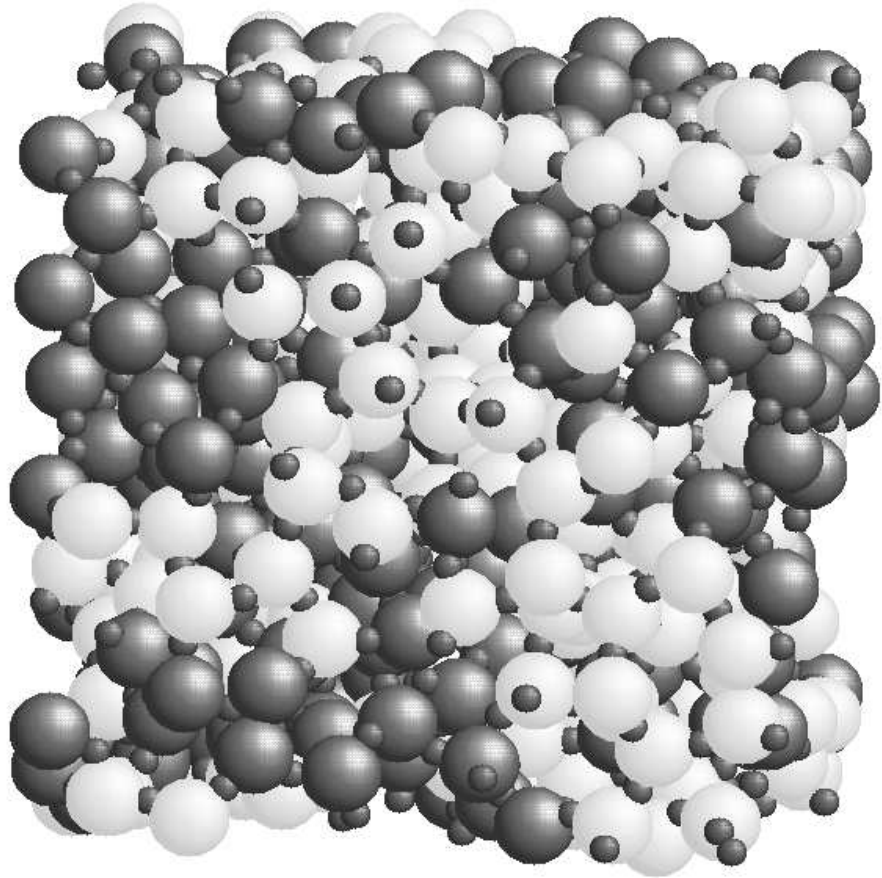,height=12cm}
\caption{Snapshot of AS2 at $T=300$~K. The big white spheres are the silicon atoms,
the big black spheres are the aluminium atoms and the small black spheres are the
oxygen atoms. Note that the size of the spheres does not correspond to the actual size 
of the atoms.}
\label{fig9}
\end{figure}

The structure of our AS2 model at $T=300$~K is illustrated by a snapshot
of the simulation box, Fig.~\ref{fig9}. (Note that the size of the
shown atoms does not correspond to their actual size.) One can clearly
see that Al rich regions are formed that percolate through the SiO$_4$
network. From this figure it is also visible that the packing of the
AlO$_4$ tetrahedra is denser due to the formation of more compact
structural units such as small--membered rings.

If our interpretation of the prepeak at $0.5$~\AA$^{-1}$ is correct
then it should be present in a pronounced way in static concentration
fluctuations. For a mixture with the two components A and B the static
concentration--concentration structure factor $S_{\rm cc}(q)$ can be
easily computed from the partial structure factors $S_{\alpha
\beta}(q)$~\cite{hansen}:
\begin{equation}
  S_{\rm cc}(q) = x_{\rm B}^2 S_{\rm AA}(q) + x_{\rm A}^2 S_{\rm BB}(q)
          - 2 x_{\rm A} x_{\rm B} S_{\rm AB}(q)
\label{eq6}
\end{equation}
with $x_{\rm A}= N_{\rm A}/N$ and $x_{\rm B}= N_{\rm B}/N$ being the
concentration of A and B particles, respectively. For our three--component
system we may also use Eq.~(\ref{eq6}) in the following way: We do not
distinguish between two of the three species such that for instance
the A species combines Si and O particles and the B species is given
by the Al particles.  For this example the functions $S_{\rm AA}(q)$,
$S_{\rm AB}(q)$, and $S_{\rm BB}(q)$ are then given by
\begin{eqnarray}
  S_{\rm AA}(q) & = & S_{\rm SiSi} + S_{\rm OO} + 2 S_{\rm SiO} \; , \nonumber \\ 
  S_{\rm AB}(q) & = & S_{\rm SiAl} + S_{\rm AlO} \; , \nonumber \\ 
  S_{\rm BB}(q) & = & S_{\rm AlAl} \; , \nonumber
\end{eqnarray}
and from these functions a structure factor $S_{\rm cc}(q)$ can be
calculated by Eq.~(\ref{eq6}) whereby $N_{\rm A}=N_{\rm Si}+N_{\rm O}$
and $N_{\rm B}=N_{\rm Al}$. The latter concentration--concentration
structure factor is nothing else than the static autocorrelation function of the
concentration density $c_{\rm Al}$ of the Al atoms~\cite{bletry,horbach03}.
In an analogous way one can calculate two other structure factors for
AS2 that correspond to the autocorrelation functions of the concentration
densities for Si and O, respectively. Thus, in terms of the concentration
densities
\begin{equation}
  c_{\alpha}({\bf q}) = \rho_{\alpha}({\bf q})
   - x_{\alpha} \left(\rho_{\rm Si}({\bf q}) 
   + \rho_{\rm Al}({\bf q}) + \rho_{\rm O}({\bf q}) \right)  
    \quad \alpha \in \{ {\rm Si, Al, O} \}
\end{equation}
with $x_\alpha = N_\alpha / N$, 
the latter three concentration--concentration structure factors are defined by
\begin{equation}
   S_{c_{\alpha} c_{\alpha}} (q) =
   \frac{1}{N} \left< c_{\alpha} ({\bf q}) c_{\alpha}(-{\bf q}) \right>  \  .
   \label{eqscc}
\end{equation}
More details on the $S_{c_{\alpha} c_{\alpha}} (q)$ can be found in
Refs.~\cite{bletry,horbach03}.

The three functions $S_{c_\alpha c_\alpha}(q)$ are shown in
Fig.~\ref{fig10} for $T=300$~K. In all three functions a pronounced peak
is seen around $q=2.72$~\AA$^{-1}$ corresponding to the length scale of
nearest Si--Si, Al--Al etc.~neighbors. This is reasonable since there
should be strong concentration fluctuations on this length scale due to
the chemical ordering in the tetrahedral network of AS2. Furthermore
we do also find a well--pronounced peak around $0.5$~\AA$^{-1}$ in
$S_{\rm c_{\rm Al} c_{\rm Al}}(q)$ and $S_{c_{\rm Si} c_{\rm Si}}(q)$
which confirms our interpretation of this prepeak. Also in accordance
with our interpretation is that no prepeak around $0.5$~\AA$^{-1}$ is
seen in $S_{c_{\rm O} c_{\rm O}}(q)$ because the O atoms are ordered in
a similar way around the cations and so in $S_{c_{\rm O} c_{\rm O}}(q)$
no distinction is made between Si rich and Al rich regions.

Of course the question arises whether it is possible to observe the prepeak
at $0.5$~\AA$^{-1}$ also in experiments, i.e. whether or not real AS2 
shows a structure on the length scale of 12.6~\AA. 
Unfortunately in experiments such as neutron scattering one does
not have access to the partial structure factors for systems like AS2
(due to the lack of appropriate isotopes) and one can only measure
a linear combination of the partial structure factors (in the case of
neutron scattering the $S_{\alpha \beta}(q)$ are weighted by the neutron
scattering lengths).  As it is shown elsewhere~\cite{winkler02}, in
such a quantity one can hardly identify the prepeak at $0.5$~\AA$^{-1}$
because at small wave--vectors the dominant contribution comes from
$S_{\mathrm{OO}}(q)$, a partial structure factor which does not show the
prepeak. So it remains a challenge to the experimentalists to develop
techniques with which one can verify the presence of the latter prepeak
in AS2.

However, in order to see how far our microscopic model is able to
reproduce the structure of real AS2, we compare now the ``reduced"
total static X--ray scattering factor of AS$2$, $q(S_{\rm X}(q)-1)$, as
obtained from our simulation to an experimental result.  $S_{\rm X}(q)$
can be calculated from the $S_{\alpha\beta}(q)$ by weighting them with
X--ray form factors:
\begin{equation}
S_{\rm X}(q)=\frac{N}{\sum_\alpha N_\alpha f_\alpha
^2(s)} \sum_{\alpha\beta}
f_\alpha(s) f_\beta(s) S_{\alpha\beta}(q)\;,
\label{eq7}
\end{equation}
with $\alpha , \beta \in\{\mathrm{Si,Al,O}\}$. The formfactors
$f_\alpha(s)$ depend on the wave--vector $q$ via $s=q/4\pi$. We have
taken the $f_\alpha(s)$ from Ref.~\cite{inttables74}. Fig.~\ref{fig11}
shows $S_{\rm X}(q)$ as calculated from our simulation by
Eq.~(\ref{eq7}) in comparison to the experimental result of Morikawa {\it et
al.}~\cite{morikawa82} for an aluminium silicate melt with 37.1~mol\%
Al$_2$O$_3$, i.e.~a composition which is slightly different from that
of AS2 (AS2 contains 33.3~mol\% Al$_2$O$_3$).  As we recognize from
Fig.~\ref{fig11} the agreement between simulation and experiment is
particularly good for $q<2.3$~\AA$^{-1}$ and also a fair agreement
is obtained for higher $q$.

Finally, we want to turn our attention to the diffusion dynamics of the
AS2 melt. The self--diffusion constant $D_\alpha$ for a particle of type
$\alpha\in\{\mathrm{Si,Al,O}\}$ can be calculated from the mean squared
displacements $\langle r^2_\alpha(t)\rangle$ via the Einstein relation:
\begin{equation}
D_\alpha = \lim_{t \to \infty}\frac{\left <r^2_\alpha(t)\right>}{6t}\,.
\label{einsteinrel}
\end{equation}
In Fig.~\ref{fig12} the three different $D_{\alpha}$ for AS2 are plotted
on a semi--logarithmic scale as a function of inverse temperature.
Also included are $D_{\rm Si}$ and $D_{\rm O}$ of pure silica from a
recent MD simulation~\cite{horb99_2}. At $T=6100$~K the $D_{\alpha}(T)$
in AS2 are very similar to those in SiO$_2$. However, upon decreasing
the temperature, the dynamics in AS2 does not slow down as rapidly
as the one of SiO$_2$. At $T=2750$~K the diffusion of all components
in AS2 is about two orders of magnitude faster than in SiO$_2$. The
self--diffusion constants in SiO$_2$ show a crossover from a power law
behavior as predicted by the mode coupling theory (MCT) of the
glass transition~\cite{mct} at high temperatures (see dashed lines in
Fig.~\ref{fig12}) to an Arrhenius behavior at low temperatures whereby
we have found the critical mode coupling temperature at 3330~K (for
more details see Refs.~\cite{horb99_2,horb01_1}). A similar analysis as
the one in SiO$_2$ for AS2 is the subject of future work.  As we see in
Fig.~\ref{fig12} the self--diffusion in AS2 provides the very interesting
case of a tetrahedral network where the dynamics of one of the cations,
aluminium, is slightly faster than that of oxygen whereas the diffusion
of the other cation, silicon, is a factor of 2 or 3 slower than that
of oxygen.

In order to quantify further the difference in the temperature dependence
of the different diffusion constants, we show in Fig.~\ref{fig13}
the ratios $D_{\rm Si}/D_{\rm O}$, $D_{\rm Si}/D_{\rm Al}$, $D_{\rm
Al}/D_{\rm O}$ for AS2 as well as $D_{\rm Si}/D_{\rm O}$ for SiO$_2$
as a function of temperature. $D_{\rm Si}/D_{\rm O}$ for SiO$_2$ varies
only from 0.65 to 0.8 in the temperature range $3400$~K~$\le T \le$
6100~K, whereas it decreases rapidly below about $3400$~K. The latter
behavior can be explained by a change in the transport mechanism around
the critical MCT temperature $T_c=3330$~K of our SiO$_2$ model (for
more details see Refs.~\cite{horb99_2,horb01_1}).  As mentioned before,
in the case of AS2 it is an open question whether one can understand
the diffusion dynamics by means of MCT. However, $D_{\rm Si}/D_{\rm O}$
and $D_{\rm Si}/D_{\rm Al}$ exhibit a very similar behavior as $D_{\rm
Si}/D_{\rm O}$ in SiO$_2$. It is remarkable that $D_{\rm Al}/D_{\rm O}$
seems to approach a constant value for temperatures below about 3000~K,
an issue that has to be clarified also in future studies.

\section{Summary and conclusions}
\label{sec4}

In this paper we have presented the results of large scale molecular
dynamics computer simulations to investigate the structure and diffusion
dynamics of the amorphous aluminium silicate melt (Al$_2$O$_3$)2(SiO$_2$)
[AS2]. The microscopic interactions were described by a simple pair
potential that has been determined by Kramer {\it et al.}~\cite{kramer}
using {\it ab initio} calculations.

It is one of the open questions in the literature on aluminium silicate
melts and glasses how the local structure around an aluminium atom
evolves in order to yield charge neutrality on a local scale. We find
in the simulations of our AS2 model that, similar to the Si atoms,
already at high temperatures most of the Al atoms are four--fold
coordinated by O atoms. However, the packing of the AlO$_4$ tetrahedra
is denser than that of the SiO$_4$ tetrahedra which is manifested
especially in the presence of 3(Al,Si) triclusters. Evidence for
such structural units in systems like AS2 has been given in NMR
experiments~\cite{poe92,meinhold93,schmuecker96,schmuecker97,schmuecker99}.
Our simulation allows a detailed description of the geometry of the
triclusters: We find that in most of them the three--fold coordinated O
atoms are members of a two--membered ring (whereby these rings contain
most likely two aluminium atoms). Furthermore, we find also a relatively
high occurrence of three--membered rings in AS2 in which most likely at
least two Al atoms participate. In contrast to the small--membered rings
in SiO$_2$, Al rings in AS2 of length two or three become slightly more
frequent if one cools down the system to low temperatures.

Similar structural features as in our AS2 model have been found
recently also in a MD simulation of pure amorphous aluminium oxide
(Al$_2$O$_3$)~\cite{guti02}. Also in that case a high number of
small--membered rings was found, in particular also a relatively high number
of two--membered rings. The geometry of the latter rings is slightly
different from that in AS2 since in Al$_2$O$_3$ (at least as far as it is
predicted in the simulation) it is not the presence of two edge--sharing
AlO$_4$ tetrahedra that is very likely but either two edge--sharing
AlO$_5$ polyhedra or an AlO$_4$ tetrahedron sharing an edge with an
AlO$_5$ unit. However, the presence of two--membered rings in a model of
Al$_2$O$_3$ and in a model of AS2 (based on completely different {\it ab
initio} calculations) may indicate that two--membered Al rings are indeed
a typical structural unit in melts and glasses containing Al$_2$O$_3$
and thus, this finding is worth to be checked in experiments.

\begin{figure}
\psfig{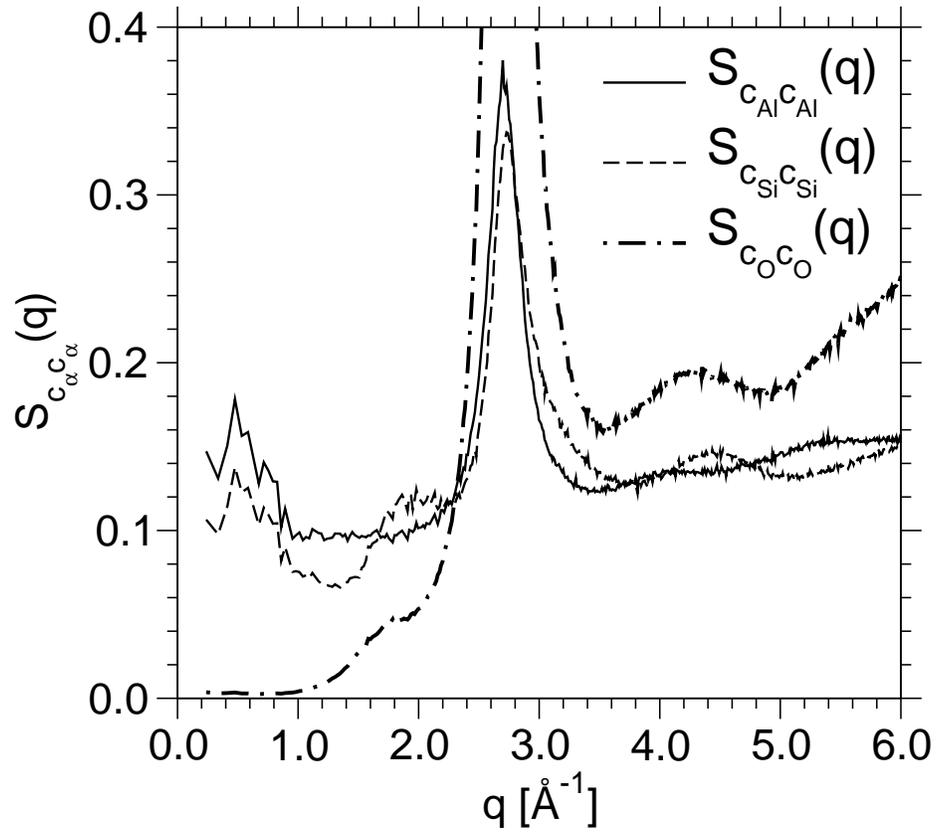}
\caption{Structure factors $S_{c_\alpha c_\alpha}(q)$ ($\alpha \in [{\rm Si, Al, O}]$)
         at $T=2300$~K. See Eq.~(\ref{eqscc}) for the definition of
        $S_{c_\alpha c_\alpha}(q)$.} 
\label{fig10}
\end{figure}

\begin{figure}
\psfig{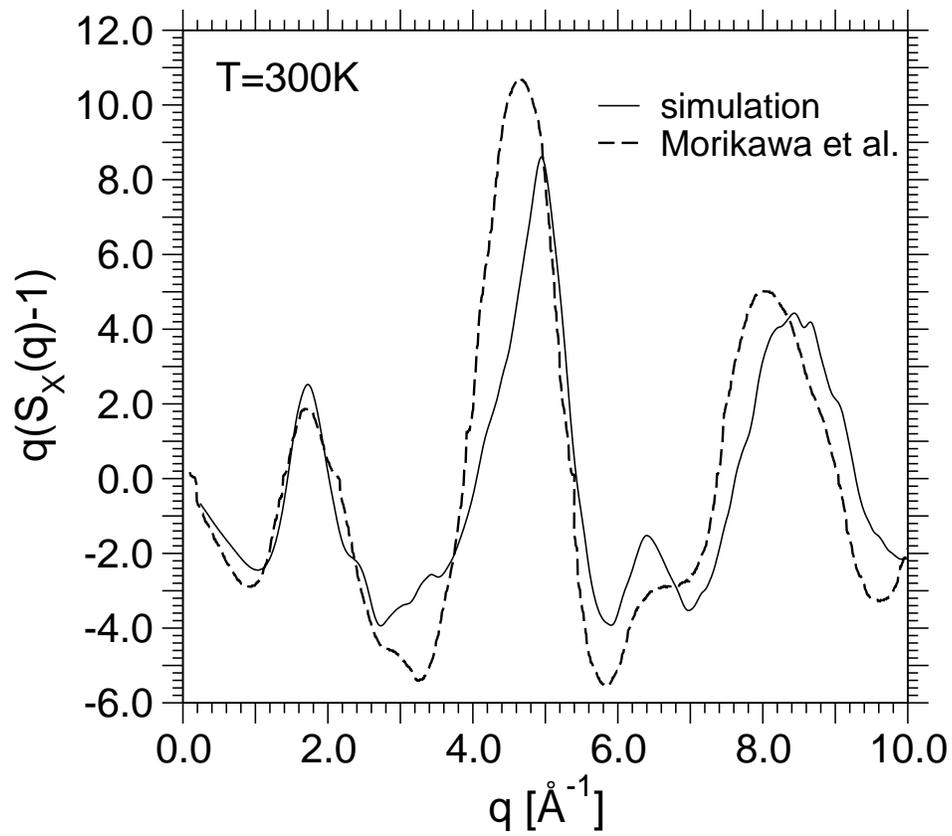}
\caption{``Reduced'' X--ray scattering factor $q(S_{\rm X}(q)-1)$ as calculated 
from the simulation using Eq.~(\ref{eq7}) (solid line) in comparison to the experimental result 
by Morikawa {\it et al.}~\cite{morikawa82} (dashed line).}
\label{fig11}
\end{figure}

\begin{figure}
\psfig{file=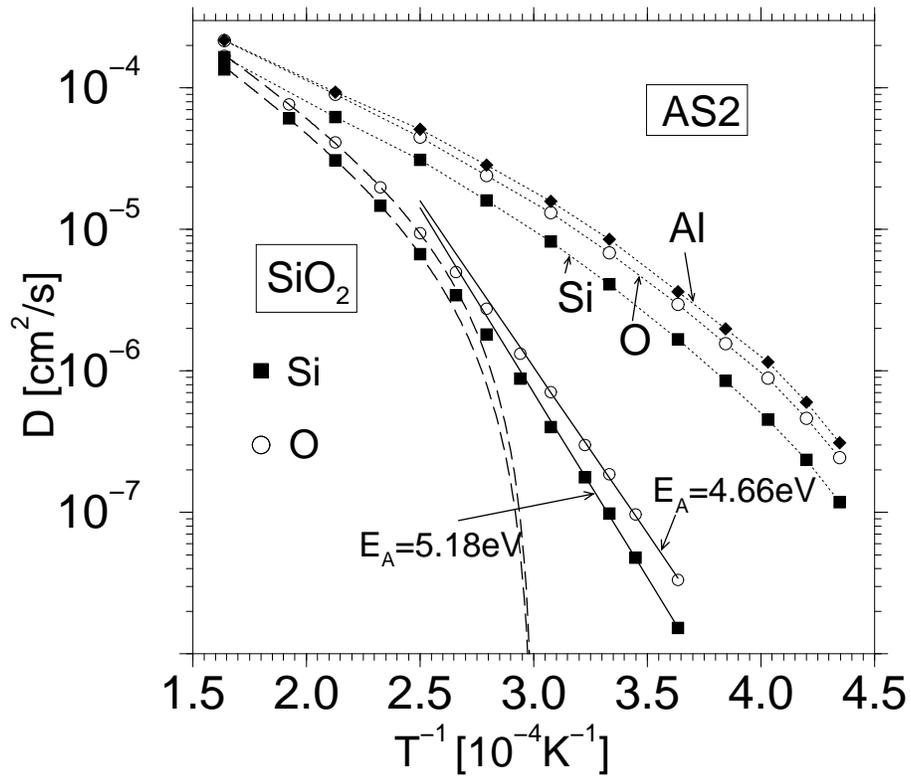,height=10cm}
\caption{Arrhenius--plot of the diffusion constants of silicon, aluminium
and oxygen: Comparison of the systems AS2 and SiO$_2$ (the data for SiO$_2$
is taken from Ref.~\cite{horb99_2}). The dashed lines are power law fits to the
SiO$_2$ data (see text for details).}
\label{fig12}
\end{figure}

\begin{figure}
\psfig{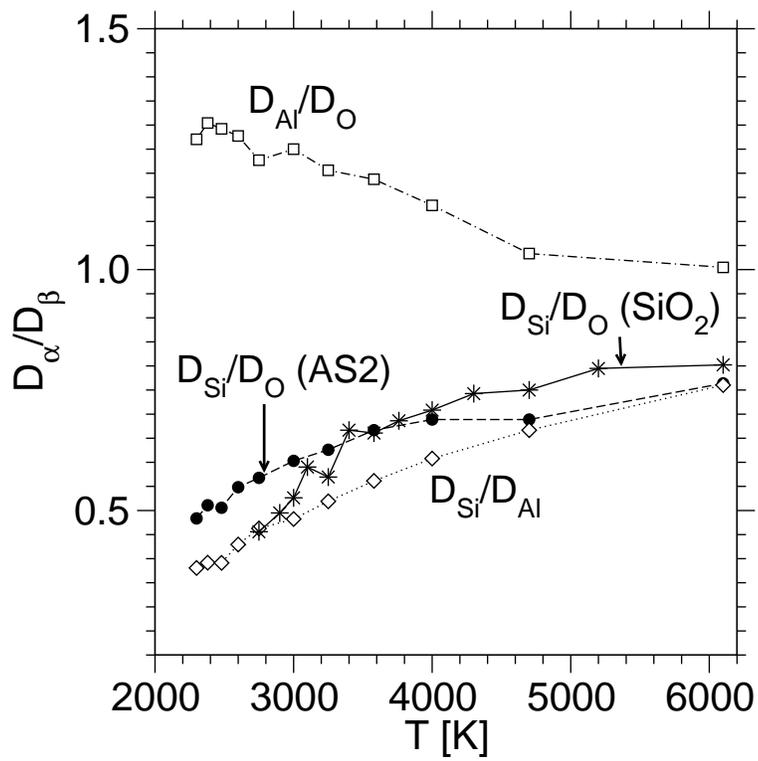}
\caption{Temperature dependence of the indicated ratios of the diffusion constants $D_{\alpha}/D_{\beta}$
         ($\alpha, \beta \in \{{\rm Si, Al, O}$\}). }
\label{fig13}
\end{figure}

The difference in the local environment of the Al and the Si atoms leads
to the formation of Al rich regions such that a network of AlO$_4$
units percolates through the SiO$_4$ structure. The characteristic
length scale of the latter structure is reflected by a prepeak in the
static structure factor at $q\approx 0.5$~\AA$^{-1}$. This prepeak can
be seen as the manifestation of a microphase separation and indeed, in
accordance with this interpretation, it is a pronounced feature in static
structure factors of concentration densities as defined by Eq.~(\ref{eq6})
(see Fig.~\ref{fig10}). The presence of a microphase separation in the melt
structure is another prediction of our simulation that could be checked
in experiments. Of course, as we have shown above, this is not a simple
task. 

It is interesting that prepeaks with a physical origin as the one
in our AS2 model have been found also for other binary silicate melts,
e.g.~in sodium silicates at $q=0.95$~\AA$^{-1}$ both in simulation and
neutron scattering or in a calcium silicate glass at $q=1.3$~\AA$^{-1}$
by means of neutron scattering~\cite{gaskell91}.

The critical point of the demixing transition in the experimental
phase diagram of Al$_2$O$_3$--SiO$_2$ is at $\approx 1920$~K for a
mixture with $\approx 30$~mol\% Al$_2$O$_3$~\cite{mcdowell69}.  Thus,
if the model used in this work exhibits a similar phase diagram as
real systems, AS2 is close to the critical composition and the lowest
temperature used in this work above the glass transition, $T=2300$~K,
is only a few hundred Kelvin above the critical temperature. In this
sense, the observed microphase separation can be seen as the precursor
of a critical unmixing transition.  In this context it is interesting
to study how the dynamics is affected by a possible interplay between
structural relaxation and a critical slowing down. It has recently been
shown that mode coupling theory (MCT) describes the slowing down due
to structural relaxation in systems like SiO$_2$ or (Na$_2$O)2(SiO$_2$)
very well~\cite{horb99_2,horb01_1,horb02_2} (this holds above and around
the critical temperature of MCT). In AS2 the case might occur where
the scenario for structural relaxation as predicted by MCT interferes with
the critical dynamics near a second order demixing transition.

In order to shed light on these guesses it is necessary to determine
the phase diagram of the model and this requires the use of techniques
that are different from Molecular Dynamics: Well--suited in this case
are Monte--Carlo simulations in the semi--grandcanonical ensemble in
conjunction with multicanonical sampling~\cite{landau_binder}. These
methods have also the advantage that they yield configurations exactly
on the coexistence line of the demixing transition which can be used
as starting configurations for Molecular Dynamics simulations to
study the microscopic structure and dynamics. The latter procedure
has recently been successfully used for a symmetrical Lennard--Jones
mixture~\cite{das03}. Therefore it can be hoped that the application
of this technique will also be useful for more complex systems like AS2
and thus will help to increase our understanding of this glass--former.

Acknowledgments:
This work was supported by SCHOTT Glas. Generous grants of computing
time on CRAY--T3E at the NIC J\"ulich and the HLRS Stuttgart are
gratefully acknowledged.

\newpage


\end{document}